\documentstyle[preprint,prb,aps]{revtex}

\begin{document}
\draft
\preprint{Impurity effect, Nov. 2000}

\title{Influence of impurity scattering on
tunneling conductance in normal metal-- $d$ -wave superconductor
junctions }

\author{Y. Tanaka and Y. Tanuma}

\address{Department of Applied Physics, Nagoya University
and CREST of the Japan
Science and Technology
Corporation (JST), Nagoya,
464-8603, Japan.}

\author{S. Kashiwaya}

\address{Electrotechnical Laboratory and CREST of the Japan
Science and Technology
Corporation (JST), Umezono, Tsukuba, Ibaraki
305-8568, Japan.}

\date{\today}
\maketitle
\begin{abstract}
Tunneling conductance spectra between
a normal metal / $d$-wave superconductor
junction under the presence of bulk impurities
in the superconductor
are studied.
The quasiclassical theory has been applied to calculate
the spatial variation of the pair potential
and the effect of impurity scattering
has been introduced
by $t$-matrix approximation.
The magnitude of a subdominant $s$-wave component at the interface
is shown to robust
against the impurity scattering while
that for  a subdominant $d_{xy}$-wave component
is largely suppressed
with the increase of the impurity scattering rate.
The zero-bias conductance peak
due to the zero-energy Andreev bound
states is significantly broadened for the
case of
Born limit impurity compared with that of
unitary limit impurity.
\end{abstract}
\par
\newpage
\narrowtext
~
Recent extensive experimental and theoretical
studies have revealed that the pair potentials of
high-$T_{\rm c}$ superconductors have
$d_{x^{2}-y^{2}}$-wave symmetry.
\cite{Scalapino,Van,Tsuei,Kashiwaya00}.
One of the most remarkable  differences of
$d_{x^{2}-y^{2}}$-wave superconductors
from conventional $s$-wave superconductors
is the presence of internal phase
of the pair potential.
As for tunneling phenomena,
the appearance of zero-bias conductance peak (ZBCP)
in tunneling conductance at [110] surface
of $d_{x^{2}-y^{2}}$-wave superconductors
reflects the sign change of effective
pair potential through
the reflection of
quasiparticle at the surface.
\cite{Hu,Tanaka1}
Qualitative features on the
orientational dependence of the
tunneling spectra in
high-$T_{c}$ superconductors have been experimentally
checked by several groups and good consistencies have been
obtained
\cite{Kashi95,Kashi96,Wei,Ig,Iguchi}.
However, detailed line shape of the conductance spectra
are not fully understood yet.
For example, the ZBCP is expected to become $\delta$-functional form
for the high-barrier limit case based on the theoretical formula,
while  experimental spectra are largely
broadened in usual cases and the origin of the broadening
is still an open problem.
\par
Possible explanations for the broadening
are the interface roughness and the impurity scattering
which may inevitably exist in the actual samples.
Several existing theories clarified the
influence of the diffuse scattering at the surface  on the
surface density of state
\cite{Matsumoto1,Yamada,Fogel,Barash97}
and on the Josephson current \cite{Barash96,Tanaka99}.
On the other hand, Poenicke, Barash, Bruder and Istyukov
(PBBI theory) studied the influence of the
bulk impurity on the surface density of states
and Josephson current \cite{Poenicke}.
They discovered that the broadening of Andreev
bound states due to unitary scattering is substantially
weaker than that in Born limit scattering.
Although PBBI theory clarified
important aspects of the
bulk impurity effect on the charge transport effect on the
$d$-wave superconductor junctions,
there still remain several open problems.
One is the detailed dependence of the conductance spectra
on transparency of the junction.
In PBBI theory, only the junctions with low transparency limit
is treated.
However,
transparency of the junctions in actual experiments
is not restricted to this limit.
Actually, it has been revealed by the previous theories that the
line shape of the tunneling conductance strongly depends  on the
transparency of the junction \cite{Kashiwaya00,Tanaka1,Kashi96}.
The other is the effect of the bulk impurity on the local inducement
of the
subdominant pair potentials
which break the time reversal symmetry.
This is important because
recent theoretical and experimental study
have shown the possibility that a subdominant
$s$-wave component or $d_{xy}$-wave component
is locally induced near the interface of
$d_{x^{2}-y^{2}}$-wave superconductor
\cite{Fogel,Sigrist95,Kuboki,Matsumoto2}.
On the other hand, the appearance of the
predicted splitting of the ZBCP's
has been detected only in several restricted experiments
and conflicting results have been reported thus far
\cite{Covington,SawaISS,Krupke}.
Therefore it is an interesting topic to
study the stability of the subdominant pair potential
under the existence of bulk impurity.
\par

In this paper, we  will calculate the tunneling conductance in
normal metal / insulator / $d_{x^{2}-y^{2}}$-wave  ($n/I/d$)
superconductor junctions using the quasiclassical formalism
of unconventional superconductors.
The self-energy is calculated in the framework of self-consistent
$t$-matrix approximation.
We concentrate on two limits,
$i.e.$ the limits of the Born scattering and the unitary scattering.
For [100] oriented junction, the obtained tunneling conductance
at zero bias voltage
is enhanced (suppressed) for high (low) transparency of the junction.
The degree of the enhancement (suppression) is much more significant for
the unitary scattering case.
While for [110] oriented junction, the broadening
of the zero bias conductance peak (ZBCP)
is much more significant for the Born scattering case
\cite{Poenicke}.
We further study the situation where a
subdominant $s$-wave component
which breaks the time reversal symmetry is induced near the interface of
$d$-wave superconductor.
The subdominant $s$-wave component  seems to be 
robust against
the presence  of the
impurity scattering.
While the
subdominant $d_{xy}$-wave component is reduced drastically
with the increase of the magnitude of the
scattering rate of impurity.
\par
The model examined here is  a two-dimensional $n/I/d$
junction within the quasiclassical formalism
\cite{Bruder,Millis}
where the pair potential has a spatial dependence
\begin{equation}
\label{A1}
\bar{\Delta}(x,\theta)
=
\left\{
\begin{array}{cc}
0,& (x \leq 0)\\
\bar{\Delta}_{R}(x,\theta), &
(x \geq 0)
\end{array}
\right.
\end{equation}
Here $\theta$ is the angle of quasiparticle trajectory
measured from the $x$ axis.
If we apply this formula to $d$-wave superconductors
including a subdominant $s$-wave and $d_{xy}$-wave
component near the interface,
$\bar{\Delta}_{R}(x,\theta)$ is decomposed into
\begin{equation}
\label{A2}
\bar{\Delta}_{R}(x,\theta)
=\Delta_{d}(x)\cos[2(\theta - \alpha)]
+ \Delta_{s}(x)
+ \Delta_{d'}(x)\sin[2(\theta - \alpha)]
\end{equation}
where $\alpha$ denotes the angle between the normal to the
interface and the $x$ axis of the crystal.
The insulator located between the normal metal and the
superconductor is modeled by a $\delta$ function.
The magnitude
of the $\delta$-function denoted as $H$ determines the
transparency of the junction $\sigma_{N}$, with
$\sigma_{N}=4\cos^{2}\theta / [Z^{2} + 4\cos^{2} \theta]$
and $Z=2mH/\hbar^{2}k_{F}$.
The effective mass $m$ and Fermi momentum $k_{F}$ are assumed to be
constant throughout the junction.
We solve the Eilenberger equation under the
existence of bulk impurity in the superconductor
\cite{Eilen,Serene}.
The tunneling conductance is obtained thorough the
coefficient of the Andreev and normal reflection
\cite{Andreev,Blonder,Zaitsev,Shelankov}. In the following,
we will concentrate on the normalized tunneling conductance
$\sigma_{T}(eV)$ \cite{Kashiwaya00,Kashi96}

\begin{equation}
\label{A8}
\sigma_{T}(eV)=
\frac{
\int^{\pi/2}_{-\pi/2} d\theta \sigma_{S}(eV,\theta) \cos\theta}
{ \int^{\pi/2}_{-\pi/2} d\theta \sigma_{N} \cos\theta}. \
\end{equation}
\begin{equation}
\sigma_{S}(eV,\theta)
= \sigma_{N}
\frac{ 1+ \sigma_{N} \left|{\eta_{R,+}(0,\theta)}\right| ^{2}
+ (\sigma_{N}-1 )\left|  \eta_{R,+}(0,\theta)
\eta_{R,-}(0,\theta)
\right|^{2} }
{ \left| {1 + (\sigma_{N} -1) \eta_{R,+}(0,\theta)
\eta_{R,-}(0,\theta)}
\right| ^{2} }.
\end{equation}
Note that $\sigma_{T}(eV)$ is expressed only by
$\eta_{R,\pm}(x,\theta)$ just at the boundary $(x=0)$
where $\eta_{R,\pm}(x,\theta)$ obeys the
following equations \cite{Nagato93},
\begin{equation}
\label{A9}
\frac{d}{dx} \eta_{R,+}(x,\theta)=
\frac{ 1}{i \hbar {v}_{F}\cos\theta }
\left[ -\Lambda_{1}(E,x,\theta_{+})
\eta_{R,+}^{2}(x,\theta)  -\Lambda_{2}(E,x,\theta_{+})
+ 2\Lambda_{3}(E,x,\theta) \eta_{R,+}(x,\theta) \right],
\end{equation}
\begin{equation}
\label{A10}
\frac{d}{dx} \eta_{R,-}
(x,\theta)=
\frac{ 1}{i \hbar v_{F}\cos\theta }
\left[ -\Lambda_{2}(E,x,\theta_{-}) \eta_{R,-}^{2}
(x,\theta)  -\Lambda_{1}(E,x,\theta_{-})
+2 \Lambda_{3}(E,x,\theta) \eta_{R,-}
(x,\theta) \right],
\end{equation}
\[
\Lambda_{1}(E,x,\theta)=
\bar{\Delta}_{R}(x,\theta_{+}) - a_{1}(E,x) + ia_{2}(E,x),
\]
\[
\Lambda_{2}(E,x,\theta)=
\bar{\Delta}_{R}^{*}(x,\theta_{+}) + a_{1}(E,x) + ia_{2}(E,x),
\]
\begin{equation}
\label{A11}
\Lambda_{3}(E,x)=E - a_{3}(E,x)
\end{equation}
with $E=eV$,
$v_{F}=k_{F}/m$, $\theta_{+}=\theta$ and $\theta_{-}=\pi-\theta$,
and $\bar{\Delta}_{R}(x,\theta_{+})$ $[\bar{\Delta}_{R}(x,\theta_{-})]$
is the effective pair potential felt by an electron [a hole] like
quasiparticle with an electron injection from the left normal metal.
The quasiparticle energy $E$ is measured from the Fermi energy.
\par
The spatial dependence of the pair potentials are determined by the following
equations \cite{ashida}
\begin{equation}
\label{B1}
\Delta_{s}(x)=g_{s}k_{B}T\sum_{\omega_{n}}
\frac{1}{2\pi}\int^{\pi/2}_{-\pi/2} d\theta'
\{
[g_{R}(\theta',x)]_{12}-[g^{+}_{R}(\theta',x)]_{12}\}
\end{equation}

\begin{equation}
\label{B2}
\Delta_{d}(x)=g_{d}k_{B}T\sum_{\omega_{n}}
\frac{1}{2\pi}\int^{\pi/2}_{-\pi/2} d\theta'
\cos[2(\theta'-\alpha)]
\{
[g_{R}(\theta',x)]_{12}-[g^{+}_{R}(\theta',x)]_{12}\}
\end{equation}

\begin{equation}
\label{B21}
\Delta_{d'}(x)=g_{d'}k_{B}T\sum_{\omega_{n}}
\frac{1}{2\pi}\int^{\pi/2}_{-\pi/2} d\theta'
\sin[2(\theta'-\alpha)]
\{
[g_{R}(\theta',x)]_{12}-[g^{+}_{R}(\theta',x)]_{12}\}
\end{equation}

\begin{equation}
\label{B22}
\lim_{x \rightarrow \infty} \Delta_{s}(x)= 0, \
\lim_{x \rightarrow \infty} \Delta_{d}(x)=\Delta_{0}, \
\lim_{x \rightarrow \infty} \Delta_{d'}(x)=0
\end{equation}
with dimensionless inter-electron coupling of the $s$-wave
$g_{s}$, $d_{xy}$-wave $g_{d'}$
and  $d_{x^{2}-y^{2}}$-wave $g_{d}$, respectively. 
The self-energy $a_{i}(E,x)$ ($i=1,3$)
describes the impurity scattering and
is given by
\begin{equation}
\label{C1}
a_{i}(i\omega_{n},x)=
\frac{ \frac{\hbar}{2\tau}
\frac{<g_{i}(\theta,x)>}{1-\sigma} }
{1 - \frac{\sigma}{1-\sigma} \sum_{i}
<g_{i}(\theta,x)>^{2} }
\end{equation}

\[
\hat{g}_{R}(\theta,x)=
g_{1}(\theta,x)\hat{\tau}_{1} +
g_{2}(\theta,x)\hat{\tau}_{2} +
g_{3}(\theta,x)\hat{\tau}_{3}
\]
with $\omega_{n}=2\pi k_{B}T(n+1/2)$.
In the above, $\hat{\tau}_{i}$ denotes the Pauli matrix and
$< \dots >$ means the average over the Fermi surface.
Here $\hbar/(2\tau)$ denotes the normal scattering rate,
while $\sigma$ measures the strength of a simple impurity potential.
The spatial dependence of the quasiclassical Green's function
$\hat{g}_{R}(\theta,x)$ is determined by
\cite{ashida}
\begin{equation}
\label{B3}
\hat{g}_{R}(\theta,x)=U_{R}(\theta,x,0)\hat{g}_{R}(\theta,0)
U_{R}^{-1}(\theta,x,0)
\end{equation}

\begin{equation}
\label{B4}
i\hbar v_{Fx}\frac{\partial}{\partial x}
U_{R}(\theta,x,0)
= -
\left(
\begin{array}{ll}
\Lambda_{3}(i\omega_{n},x) & \Lambda_{1}(i\omega_{n},x,\theta_{+}) \\
-\Lambda_{2}(i\omega_{n},x,\theta_{+}) & -\Lambda_{3}(i\omega_{n},x)
\end{array}
\right)
U_{R}(\theta,x,0),
\end{equation}
with  $U_{R}(\theta,0,0)=1$.
In the actual numerical calculations, $\eta_{R,\pm}(x,\theta)$
is calculated from Eqs. (\ref{A9}) to (\ref{A10}).
Since $\hat{g}_{R}(\theta,0)$ is expressed by
$\eta_{R,\pm}(\theta,0)$, $\hat{g}_{R}(\theta,x)$ is obtained
using Eqs. (\ref{B3}) to (\ref{B4}).
Subsequently, the spatial dependence of the pair potentials
$\Delta_{d}(x)$, $\Delta_{d'}(x)$ and
$\Delta_{s}(x)$ are calculated by
Eqs. (\ref{B1}) to (\ref{B22}).
The self-energies $a_{i}(i\omega_{n},x)$ are
calculated by Eq. (\ref{C1}).
To get self-consistently determined pair potentials
and self-energies due to the impurity scattering
this process is repeated until enough convergence is obtained.
In the following calculation, the two important parameters
characterizing the impurity scattering
are $\frac{\hbar}{2\tau \Delta_{c0}}$ and $\sigma$ where
$\Delta_{c0}$ denotes the magnitude of the bulk pair potential
without the existence of impurity scattering.
The magnitude of $\Delta_{0}$ 
depends on $\frac{\hbar}{2\tau \Delta_{c0}}$ and $\sigma$. 
\par
First, let us consider the case where
neither $\Delta_{s}(x)$ nor $\Delta_{d'}(x)$ present.
One of the typical example is the [100] oriented junction,
$i.e.$, $\alpha=0$ case,
where  both the transmitted electron like and hole like quasiparticle
feel the same sign of the pair potentials.
For high transparent junctions
$i.e.$ $Z=0$ [see Fig. 1(a)],
only the Andreev scattering takes place at the interface and the
resulting conductance shows zero-bias enhancement due to
the Andreev reflction without impurity.
With the introduction of the impurity scattering
the height of the peak is suppressed.
The degree of this suppression is much more significant for
unitary scattering case [see curve $c$ in Fig. 1 (a)].
On the other hand, for larger magnitude of
$Z$ [ see Fig. 1(b)],  $\sigma_{T}(eV)$ has a gap
like structure without impurity.
The magnitude of $\sigma_{T}(0)$ is enhanced with the
increase of $Z$. The degree of this enhancement is
much more significant in the unitary scattering case.
In order to understand these line shapes of the tunneling conductances,
we will look at $eV=0$,
where  $\eta_{R,\pm}(0,\theta) =\pm i$ is satisfied
without impurity scattering.
With the introduction of impurity potential,
the magnitude of $\eta_{R,\pm}(0,\theta)$
is suppressed and we can denote
$\eta_{R,\pm}(0,\theta) = \pm i \exp(-\gamma_{0})$.
Here, $\gamma_{0}$ is a function with positive value
and
enhances with the increase of the  magnitude of $a_{3}(0,0)$,
$i.e.$, the magnitude of $\frac{\hbar}{2\tau \Delta_{c0}}$.
For $\alpha=0$, since
$\eta_{R}(0,\theta_{+}) = \pm i \exp(-\gamma_{0})$ and
$\eta_{R}(0,\theta_{-}) = \pm i \exp(-\gamma_{0})$
are satisfied,
the resulting
$\bar{\sigma}_{S}(0,\theta)$
is $ [1 + \exp(-2\gamma_{0})]$  and
$[1 - \exp(-2\gamma_{0}) ]/[1 + \exp(-2\gamma_{0})]$ for
$Z=0$ and sufficiently
larger magnitude of $Z$, respectively.
Since the magnitude of $a_{3}(0,0)$ is enhanced for the
unitary scattering case, the resulting $\gamma_{0}$ also
enhances.
Consequently, the magnitude of $\bar{\sigma}_{S}(0,\theta)$ is
much more enhanced for the Born  scattering case for $Z=0$,
while the opposite situation happens for sufficiently larger value of $Z$.
\par
Next, let us look at the $\alpha=\pi/4$ case,
where $\sigma_{T}(eV)$ has a
zero bias conductance peak
(ZBCP) with small magnitude of
$\sigma_{N}$ \cite{Tanaka1,Kashi95,Kashi96}.
For $Z=0$, almost the same
result as that in Fig. 1(a) is obtained.
This is because well defined bound states do not exist
in the absence of the barrier potential.
With the increase of the magnitude of $Z$, the ZBCP shows up
due to Andreev bound state formation
without impurity \cite{Tanaka1,Kashi95,Kashi96}.
As shown in Fig. 2, the height of the ZBCP is suppressed with the
introduction of the impurity scattering.
In the Born limit (see curve $b$ in Fig. 2),
the reduction of the amplitude of the peak height
and the broadening of the  peak width is most significant.
To understand this feature, here, we focus on  $eV=0$,
where
$\eta_{R,+}(0,\theta) = i \exp(-\gamma_{0})$ and
$\eta_{R,-}(0,\theta) = -i \exp(-\gamma_{0})$
are satisfied.
The resulting
$\bar{\sigma}_{S}(0,\theta)$ is
$[1 + \exp(-2\gamma_{0}) ]/[1 - \exp(-2\gamma_{0})]$ for sufficiently
large magnitude of $Z$.
It should be remarked that the magnitude of
$a_{3}(0,0)$ is much more enhanced for Born scattering case
due to the formation Andreev bound state at the interface
\cite{Poenicke}
and the resulting $\gamma_{0}$ is much more enhanced.
Consequently, the amplitude of $\bar{\sigma}_{S}(0,\theta)$, and
$\sigma_{T}(0)$ in the unitary limit is larger than that in the
Born limit. \par
In Fig. 3, the width of ZBCP $W$
is plotted as a function of
scattering rate $\hbar/(2\tau \Delta_{c0})$ in the Born limit.
For small value of $Z$, $W$ is
insensitive with the increase of  $\hbar/(2\tau \Delta_{c0})$.
However, with the increase of the amplitude of $Z$,
$W$ becomes a monotonically increasing function of
$\hbar/(2\tau \Delta_{c0})$. For sufficiently larger value of
$Z$,  $W$ is roughly proportional to the inverse of
$\sqrt{\hbar/(2\tau \Delta_{c0}) }$ as predicted by  PBBI theory
\cite{Poenicke}.

%%%%%%%%%%%%%%%%%%%%%%%%%%%%%%%%%%%%%%%%%%%%%%%%%%%%%%%
% d +is state
%%%%%%%%%%%%%%%%%%%%%%%%%%%%%%%%%%%%%%%%%%%%%%%
Under the presence of the ZBCP, since the quasiparticle
density of states at the zero energy near the interface are enhanced,
a subdominant $s$-wave ($d_{xy}$-wave)
component of the pair potential
$\Delta_{s}(x)$ ($\Delta_{d'}(x)$ )
can be induced near the interface,
when a finite $s$-wave ($d_{xy}$-wave) pairing interaction strength exists,
even though the bulk symmetry
remains pure $d_{x^{2}-y{2}}$-wave
\cite{Fogel,Sigrist95,Kuboki,Matsumoto2,Covington}.
Since the phase difference of the
$d_{x^2-y^2}$-wave and $s$-wave components is not always a multiple of $\pi$,
the mixed state breaks the time-reversal symmetry
\cite{Fogel,Matsumoto2}.
It is an interesting problem to clarify the stability
of this broken time reversal symmetry  state (BTRSS).
Here, we study two cases;
(a) a subdominant
$s$-wave component is induced near the interface, $i.e.$,
$\Delta_{s}(x) \neq 0$, $\Delta_{d'}(x)=0$ [see Fig. 4(a)],
(b) a subdominant
$d_{xy}$-wave component is induced near the interface, $i.e.$,
$\Delta_{d'}(x) \neq 0$, $\Delta_{s}(x)=0$ [see Fig. 4(b)].
Here,  the transition temperature of the $s$-wave  component alone
$T_{s}$ and  that of $d_{xy}$-wave component
$T_{d'}$ are chosen as
$T_{s}=0.3T_{d}$ and $T_{d'}=0.3T_{d}$,
where $T_{d}$ is the transition temperature of $d_{x^{2}-y^{2}}$
pair potential without any other components.
The transition temperature
$T_{s}$, $T_{d'}$ and $T_{d}$ directly correspond to the
magnitude of attractive inter-electron coupling
$g_{s}$, $g_{d'}$ and $g_{d}$, respectively.
In Fig. 4, the relative magnitude of
the subdominant $s$-wave or $d_{xy}$ component
at the interface, $i.e.$,
$\mid {\rm Imag} \Delta_{s}(0) \mid/\Delta_{0}$
or $\mid {\rm Imag} \Delta_{d'}(0) \mid/\Delta_{0}$ 
is plotted as a function of $\hbar/(2\tau \Delta_{c0})$ 
both in the Born and unitary limit.
The magnitude of $\mid {\rm Imag} \Delta_{s}(0) \mid/\Delta_{0}$ is
insensitive with the change of the value of
$\hbar/(2\tau \Delta_{c0})$, 
while that for $\mid {\rm Imag} \Delta_{d'}(0) \mid/\Delta_{0}$
is reduced with the increase of the magnitude of
$\frac{\hbar}{2\tau \Delta_{c0}}$.
The resulting $\sigma_{T}(eV)$ is plotted in Fig. 5 for
$\frac{\hbar}{2\tau \Delta_{c0}}=0.1$ both in the Born limit
and unitary scattering case.
In the case where subdominant $s$-wave component is induced at the interface,
the resulting $\sigma_{T}(eV)$ has a ZBCP splitting independent of
the introduction of the impurity scattering.
On the other hand, the ZBCP is recovered with the introduction of
impurity scattering  when subdominant pairing is
$d_{xy}$-wave (see Fig. 5(b)).
In this case, the onset of the ZBCP splitting depends on the magnitude of
$\frac{\hbar}{2\tau \Delta_{c0}}$.
 From these behaviors, we can distinguish
the symmetry of the subdominant pair potential.
\par
In this paper, tunneling conductance between
normal metal / $d$-wave superconductor
junctions is studied under
the presence of bulk impurities in superconductors.
We use quasiclassical theory and include impurity scattering
by $t$-matrix approximation, and
the spatial variation
of the pair potentials and self-energy are determined self-consistently.
For [100] oriented junction, obtained tunneling conductance
at zero bias voltage
is enhanced (suppressed) for high (low) transparency of the junction.
The degree of the enhancement (suppression) is much more significant for
the unitary scattering case.
While for [110] oriented low transparent junctions, the broadening
of the zero bias conductance peak (ZBCP)
is much more significant for the Born scattering case
\cite{Poenicke}.
We further study the situation where a
subdominant component
which breaks the time reversal symmetry is induced near the interface of
$d$-wave superconductor
\cite{Fogel,Sigrist95,Kuboki,Matsumoto2,Covington}.
The subdominant $s$-wave component is robust against the introduction of the
impurity scattering. While the
subdominant $d_{xy}$-wave component is reduced drastically
with the increase of the magnitude of the
scattering rate of impurity.
We will comment on the relation
between present results with experimental ones.
In the cases of high-$T_c$ superconductors,
impurities intrinsically exist in real samples.
For example, the enhancement of zero-bias conductance is
reported in c-axis observation of STM experiments \cite{Hudson}.
We believe that the observed features
(especially the enhancement of conductance)
in this experiment are consistent with
the  result for the Born limit case.
On the other hand, in the cases of in-plane tunneling,
only a few papers
reported the  splitting of ZBCP in YBCO junctions\cite{Covington,Krupke}
and the others report the absence of the splitting (for example, 
\cite{SawaISS}).
If the appearance and disappearance of the splitting is
simply determined by
the Born impurity concentration in samples,
the promising subdominant component is concluded to be $d_{xy}$-wave.
However, from the microscopic calculation using the
$t-J$ model, the  subdominant
$d_{xy}$-wave is hard to be realized while the
subdominant  $s$-wave component
is much more plausible \cite{Tanuma98,Tanuma99}.
In this sence, we should perform the tunneling experiments
with controlled impurity inclusion in the
high-T$_{c}$ cuprates.
Also we must reveal the origin of the attractive
potential which induces subdominant
$d_{xy}$-wave component in order to overcome this contradiction.
Furthermore, it should be pointed out that
the magnetically active interface also
induces the ZBCP splitting \cite{Kashiwaya99,Zhu99}.
To resolve the origin  of the ZBCP splitting
in the actual  junctions,
we must study much more about microscopic electronic properties
about the interface of high-T$_{c}$ cuprates. \par
%%%%%%%%%%%%%%%%%%%%%%%%%%%%%%%%%%%%%%%%%%%%%%%%%%%5
%     Future and open problems
%%%%%%%%%%%%%%%%%%%%%%%%%%%%%%%%%%%%%%%%%%%%%%%%%%%%%
Throughout this paper, the roughness of the interface
is not taken into account. Since
the roughness of the interface suppresses the magnitude of
the ZBCP \cite{Yamada},
the low voltage behavior of the
$\sigma_{T}(eV)$  is also expected to be influenced.
It is an interesting  future problem to calculate
$\sigma_{T}(eV)$  both under the existence of
surface roughness and the bulk impurity.
In the present paper, the effect of bulk impurity is
studied where the spatial dependence of
$\tau$ is not taken into account.
In the actual sample, it is expected that the
value of $\tau$  near the interface is much more smaller than that
in the bulk.
The influence of Andreev bound state on the
Josephson effect in $d$-wave superconductors 
is also an interesting 
issue \cite{Barash96,Poenicke,Tanaka2,Tanaka3}.
To resolve the bulk impurity effect on the Josephson current
in the junctions with sufficient transparency is an interesting future
problem since the experimentally accessible junction is not always
in the tunneling limit \cite{Arie}.
\par
\vspace{1.0cm}
This work was partially supported by a Grant-in-aid for Scientific Research
from the Ministry of Education, Science, Sports and Culture.
\par

\newpage
\noindent
Figure Captions \par
\noindent
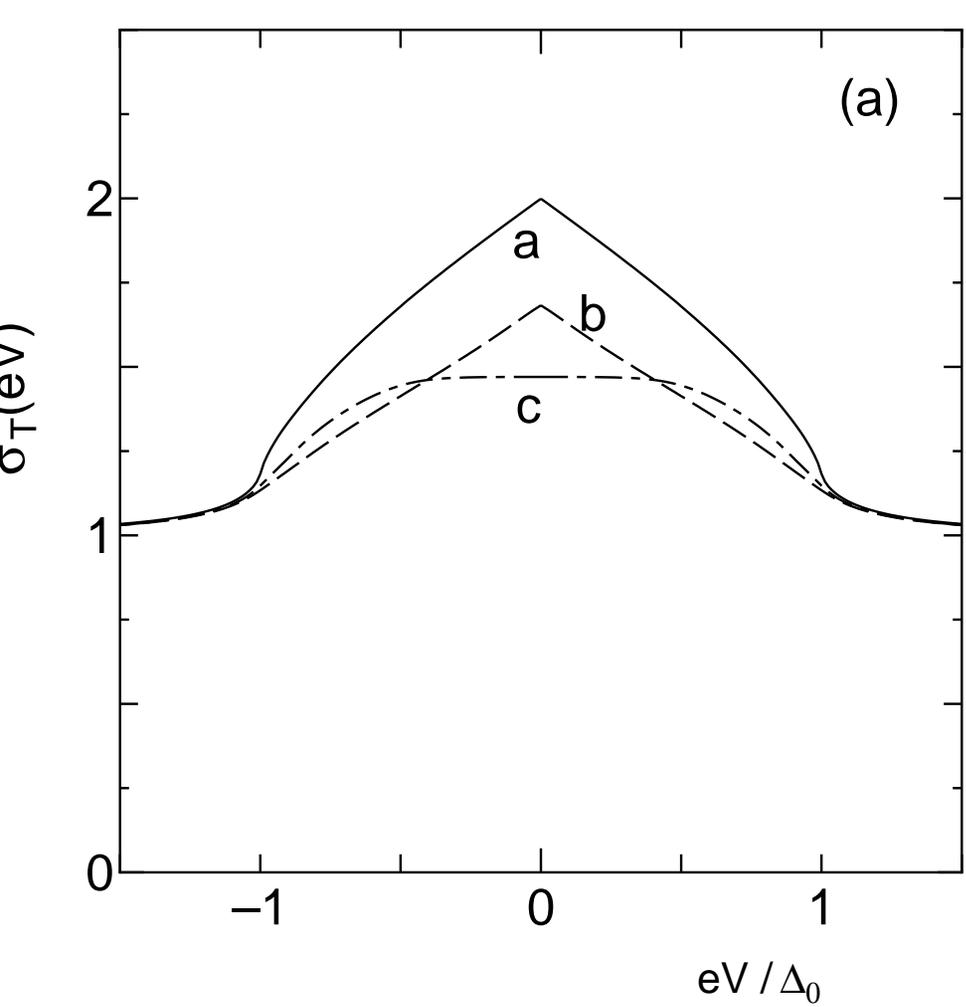
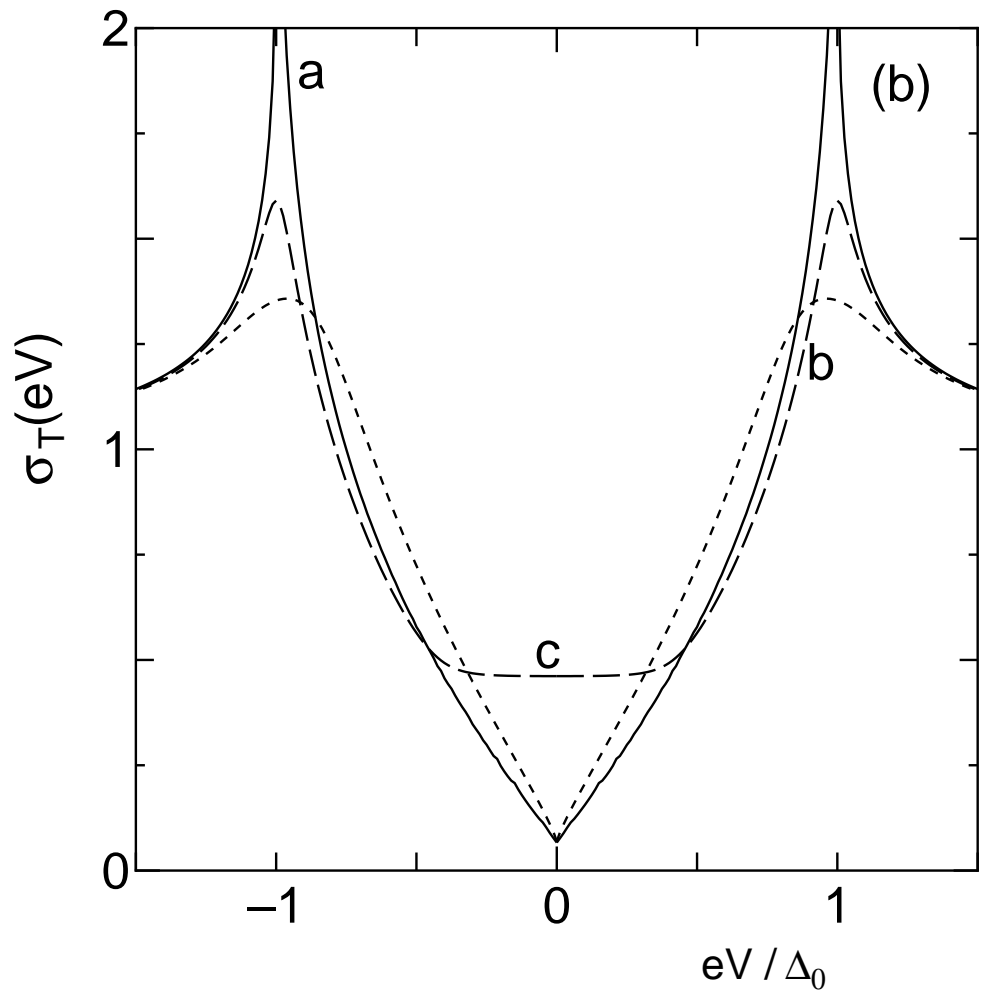
Fig. 1. $\sigma_{T}(eV)$
for $n/I/d$ junction with   $\alpha=0$.
(a) $Z=0$, (b)$Z=5$.
a: clean lmit, $\frac{\hbar}{2\tau \Delta_{c0}}=0$,
b: Born limit, $\frac{\hbar}{2\tau \Delta_{c0}}=0.1$, $\sigma=0$,
c: almost unitary limit, $\frac{\hbar}{2\tau \Delta_{c0}}=0.1$,
$\sigma=0.99$. \par
\noindent
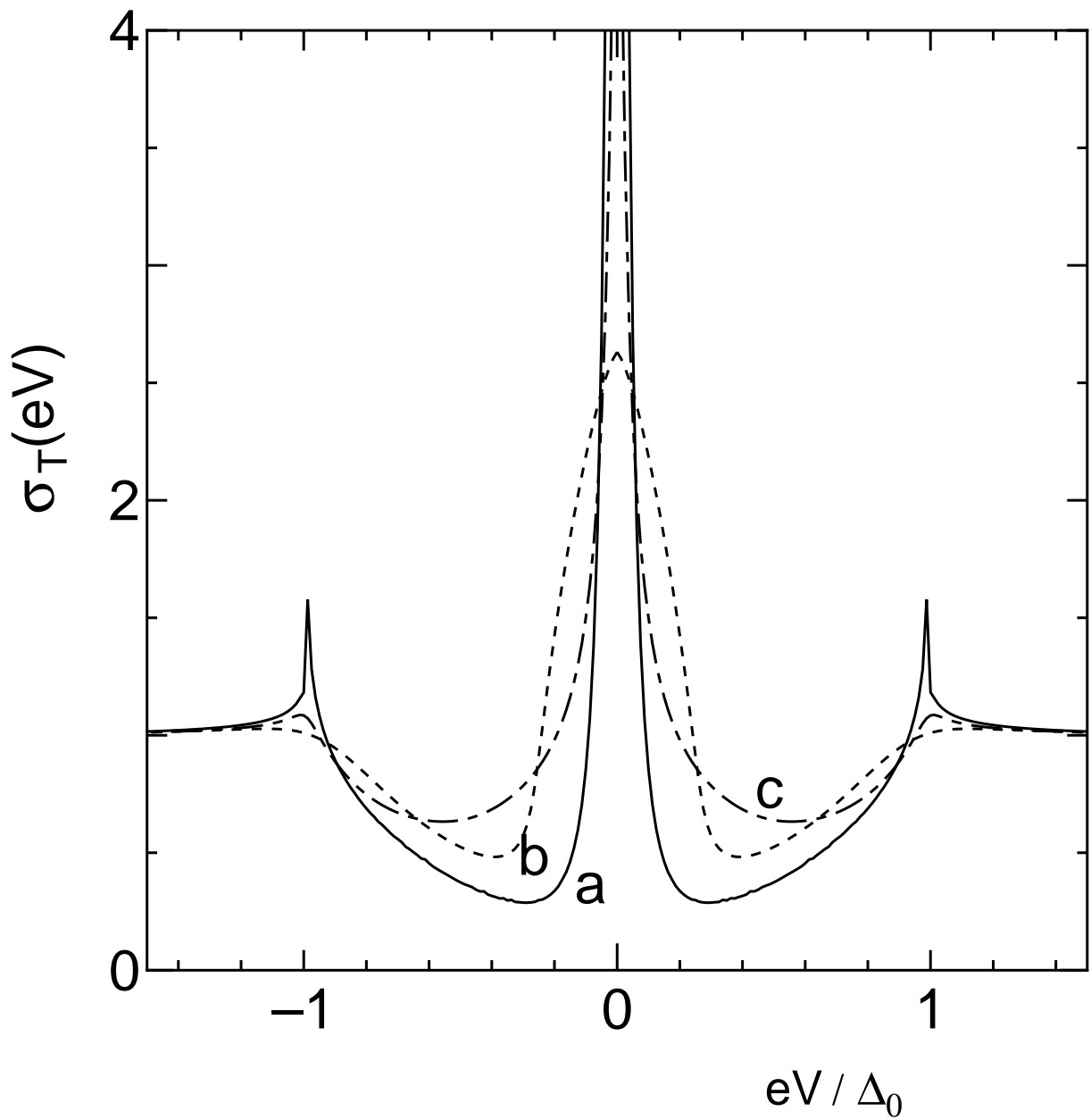
Fig. 2. $\sigma_{T}(eV)$
for $n/I/d$ junction with   $\alpha=\pi/4$
and $Z=5$.
a: clean lmit, $\frac{\hbar}{2\tau \Delta_{c0}}=0$,
b: Born limit, $\frac{\hbar}{2\tau \Delta_{c0}}=0.1$, $\sigma=0$,
c: almost unitary limit, $\frac{\hbar}{2\tau \Delta_{c0}}=0.1$,
$\sigma=0.99$. \par

\noindent
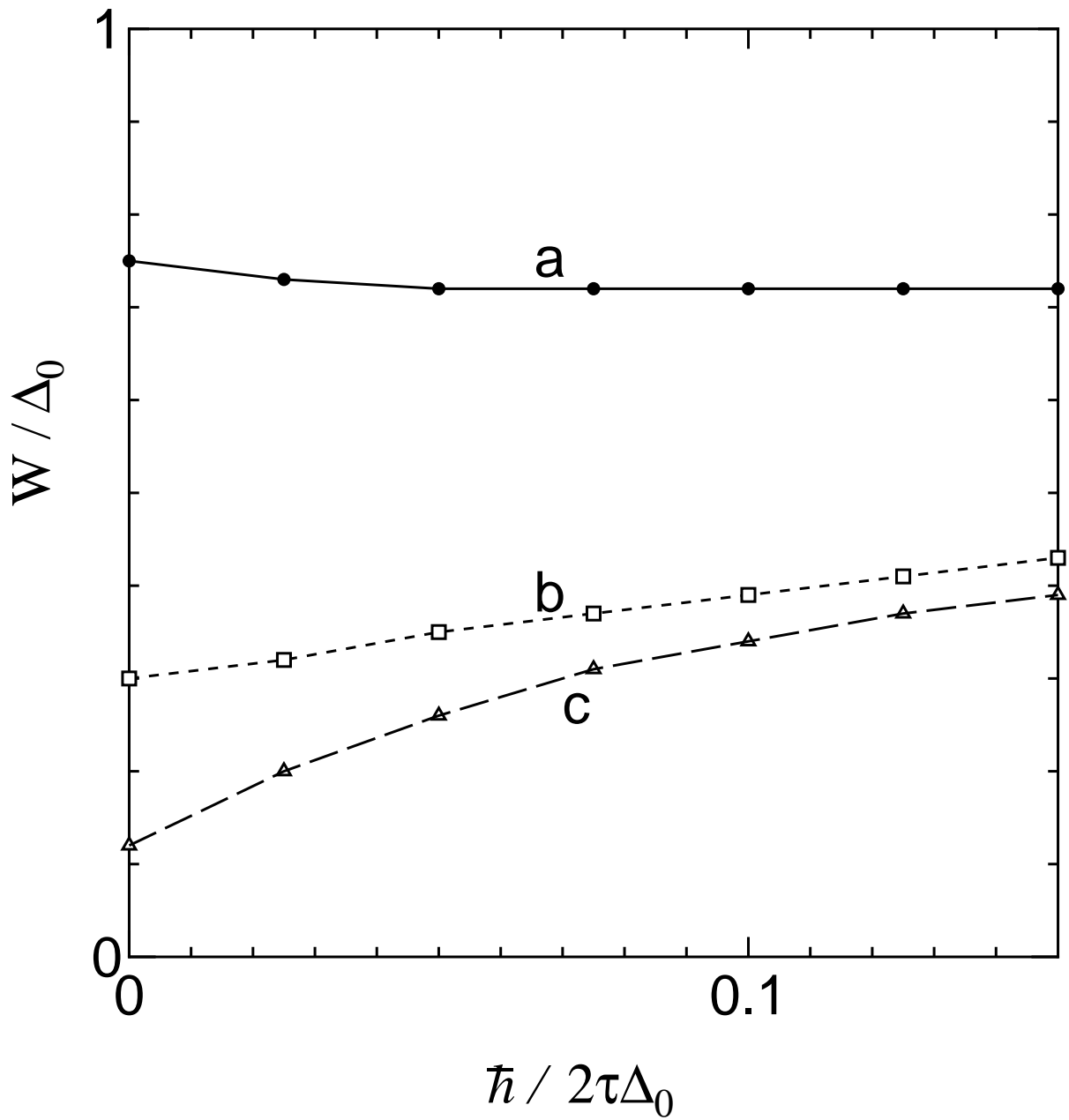
Fig. 3. The width of ZBCP $W/\Delta_{c0}$
for $n/I/d$ junction with   $\alpha=\pi/4$
is plotted as a function of
$\frac{\hbar}{2\tau \Delta_{c0}}$.
a: $Z=1$, b:$Z=5$, and c: $Z=20$.  \par
\noindent
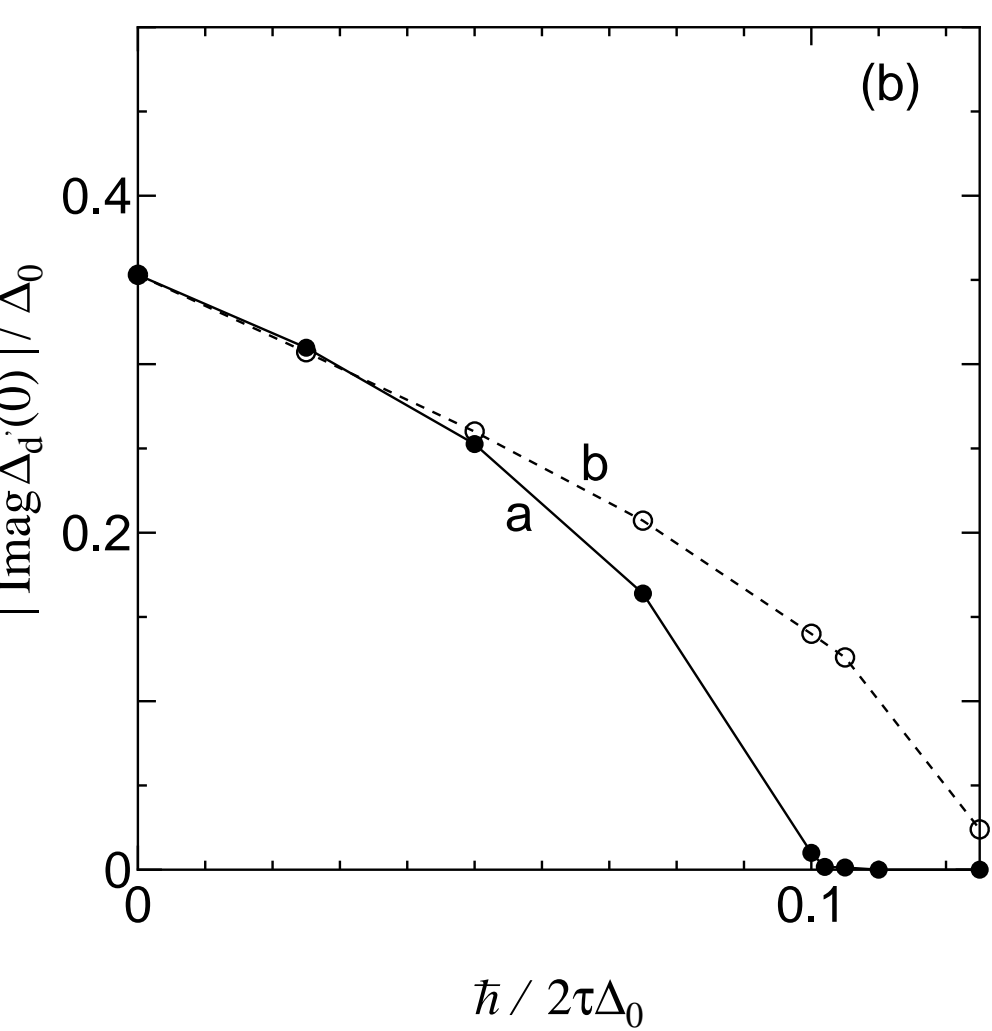
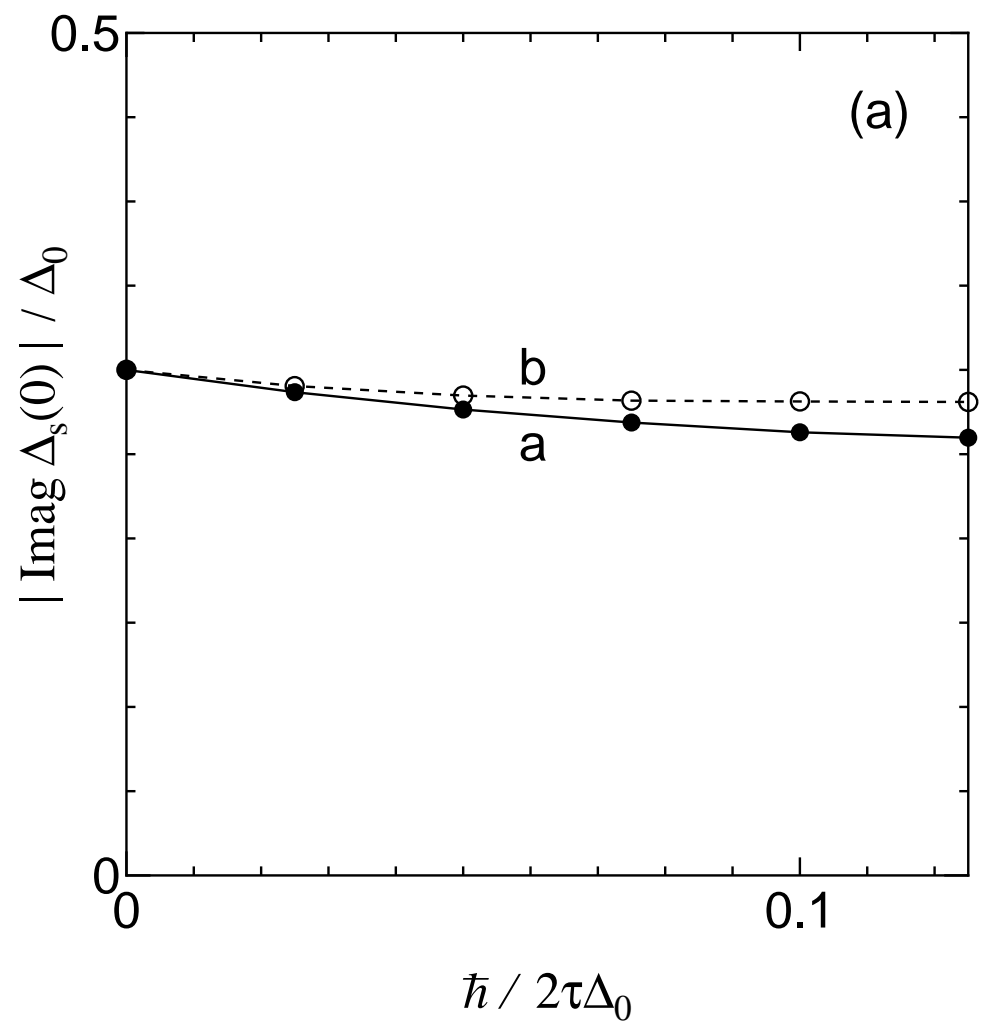
Fig. 4. The relative amplitude of the pair potential of
a subdominant $s$-wave component
and $d_{xy}$-wave component
for $n/I/d$ junction with   $\alpha=\pi/4$ and $Z=5$
is plotted as a function of
scattering strength of the impurity 
$\frac{\hbar}{2\tau \Delta_{c0}}$.
(a) subdominant $s$-wave component,
$\mid {\rm Imag} \Delta_{s}(0)\mid/\Delta_{0}$.
(b) subdominant $d_{xy}$-wave component,
$\mid {\rm Imag} \Delta_{d'}(0) \mid/\Delta_{0}$.
a: Born limit,  $\sigma=0$,
b: almost unitary limit, $\sigma=0.99$. \par

\noindent
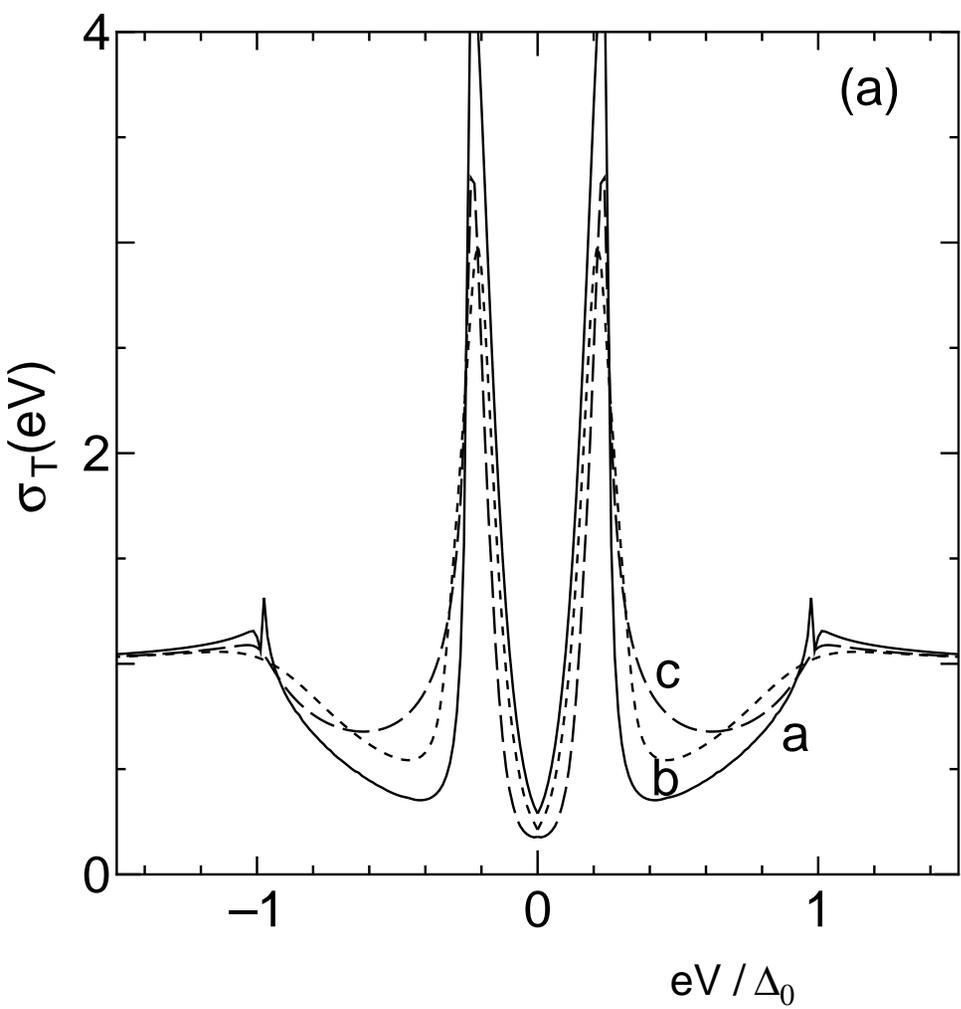
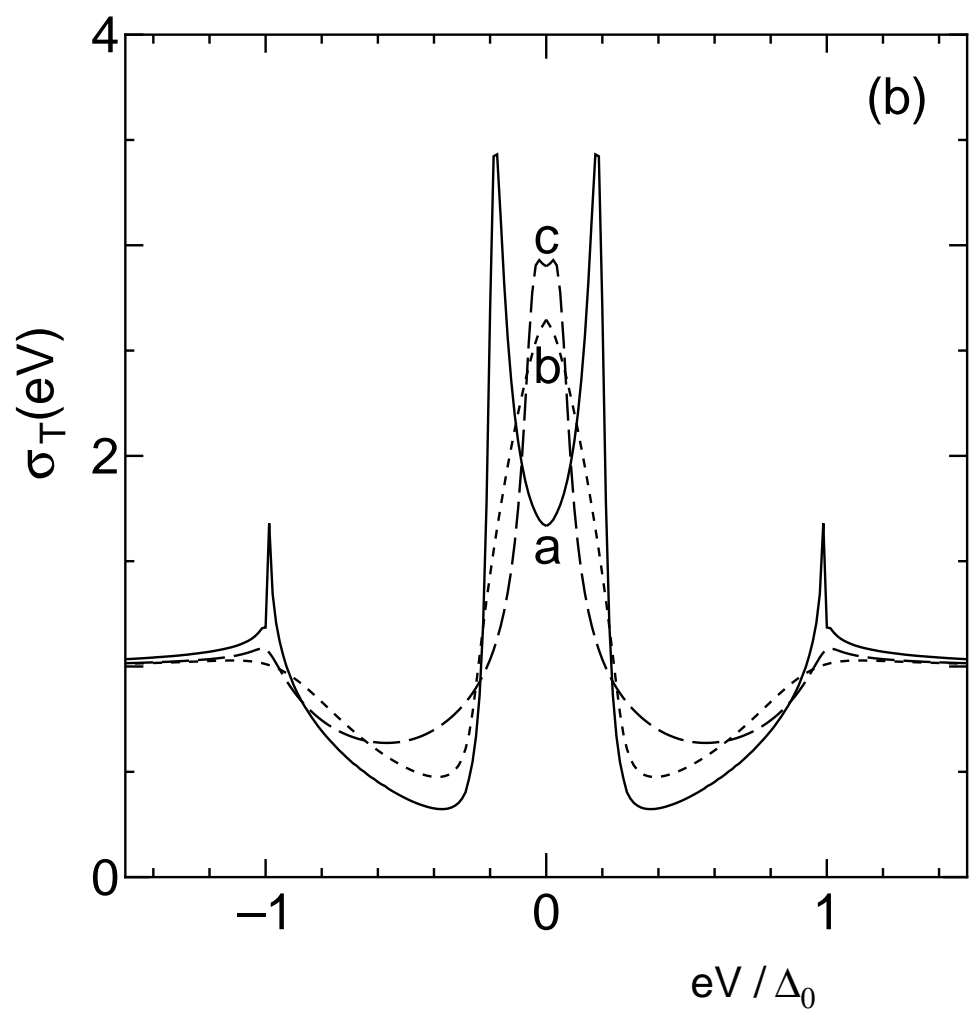
Fig. 5. $\sigma_{T}(eV)$
for $n/I/d$ junction with   $\alpha=\pi/4$ and $Z=5$
under the existence of  a subdominant $s$-wave component [(a)]
and $d_{xy}$-wave component [(b)].
a: clean lmit, $\frac{\hbar}{2\tau \Delta_{c0}}=0$,
b: Born limit, $\frac{\hbar}{2\tau \Delta_{c0}}=0.1$, $\sigma=0$,
c: almost unitary limit, $\frac{\hbar}{2\tau \Delta_{c0}}=0.1$,
$\sigma=0.99$. \par

\end{document}